\documentstyle{article}

\font\ss=cmss10
\font\scriptrm=cmr8

\def\D{\displaystyle}
\def\HHH#1{\vphantom{\vrule width 0.1pt height #1pt depth 0pt}}

\def\onlinecite#1{\cite{#1}}

\def\hr{\hbox{\bf R}}

\def\tr{\hbox{\rm tr}}
\def\re{\mathop{\hbox{\,\rm Re}}\nolimits}
\def\im{\mathop{\hbox{\,\rm Im}}\nolimits}

\def\d{\hbox{\rm d}}
\def\res{\mathop{\hbox{\rm Res}}\limits}
\def\diag{\hbox{\rm diag}}

\font\eightrm=cmr8
\newif\ifinexp \inexpfalse
\def\I{\ifinexp \hbox{\eightrm i} \else \hbox{\rm i} \fi}

\def\EA#1#2#3{{\cal E}^{(#1)}_{#2#3}}

\def\EE#1#2{E^{(#1)}_{#2}}
\def\EEE#1#2{\widetilde E^{(#1)}_{#2}}
\def\EF#1#2{F^{(#1)}_{#2}}

\def\matrix#1{\begin{array}{#1}}
\def\endmatrix{\end{array}}
\def\pmatrix#1{\left(\begin{array}{#1}}
\def\endpmatrix{\end{array}\right)}

\def\aligned#1{\begin{array}{#1}}
\def\endaligned{\end{array}}

\newtheorem{theorem}{\bf Theorem}
\newtheorem{lemma}{\bf Lemma}

\newtheorem{remark}{\sl Remark}

\title
{Liouville integrability of the finite dimensional Hamiltonian
systems\\ derived from principal chiral field}

\author
{\small Zixiang Zhou\\ \small Institute of Mathematics, Fudan
University, Shanghai 200433, People's Republic of China\\
\small E-mail: zxzhou@guomai.sh.cn}
\date{}

\begin{document}
\maketitle

\begin{abstract}
For finite dimensional Hamiltonian systems derived from 1+1
dimensional integrable systems, if they have Lax representations,
then the Lax operator creates a set of conserved integrals. When
these conserved integrals are in involution, it is believed quite
popularly that there will be enough functionally independent ones
among them to guarantee the Liouville integrability of the
Hamiltonian systems, at least for those derived from physical
problems. In this paper, we give a counterexample based on the
$U(2)$ principal chiral field. It is proved that the finite
dimensional Hamiltonian systems derived from the $U(2)$ principal
chiral field are Liouville integrable. Moreover, their Lax
operator gives a set of involutive conserved integrals, but they
are not enough to guarantee the integrability of the Hamiltonian systems.
\end{abstract}


\section{Introduction}

For many 1+1 dimensional integrable systems, the nonlinearization
method can be applied to get finite dimensional (1+0 dimensional)
Hamiltonian systems \cite{bib:Cao}. Usually these Hamiltonian
systems have Lax representations so that the involutive conserved
integrals can be obtained. In this way the original nonlinear
partial differential equations are changed to systems of
nonlinear ordinary differential equations
\cite{bib:Caomag,bib:CaoKP,bib:CaoB,bib:Ma,bib:Ragnisco1,bib:Ragnisco,bib:ZRG}.
Many interesting exact solutions, especially quasi-periodic
solutions were obtained in this way.

For a finite dimensional Hamiltonian system, if it can be written
in the Lax form as
\begin{equation}
   \frac{\d}{\d t}L(\lambda)=[M(\lambda),L(\lambda)],
\end{equation}
then the conserved integrals are easily derived from the
coefficients of $\tr(L^k(\lambda))$'s $(k\ge 1)$ when they are
expanded as Laurent series of $\lambda$. Usually the number of
these coefficients are infinite. It is believed quite popularly
that when these conserved integrals are in involution, there will
be enough functionally independent ones among them to guarantee
the Liouville integrability of the Hamiltonian systems. Indeed,
this is the case for most known physically interested systems,
such as the equations in AKNS system, Kaup-Newell system and many
other examples including those derived from 2+1 dimensional
integrable systems
\cite{bib:CaoKP,bib:CaoB,bib:Ragnisco,bib:MaZhou,bib:Zeng,bib:ZMZ}.

However, we will give a counterexample in this paper to show that
this is not always true.

This counterexample is based on a well-known physical model ---
the $U(n)$ principal chiral field (or mathematically, the
harmonic map from $\hr^{1,1}$ to $U(n)$)
\cite{bib:Prasad,bib:Uh,bib:GH,bib:Guestbook,bib:Woodbook}. In
this paper, the equation of $U(n)$ principal chiral field can be
first reduced to a set of Hamiltonian systems by the standard
procedure of the nonlinearization method. This will be done in
Sec.~\ref{sect:hm12} and Sec.~\ref{sect:Laxop}. Then, in
Sec.~\ref{sect:depend}, we show that there are not enough
conserved integrals in those given by $\tr(L^k(\lambda))$'s to
guarantee the Liouville integrability of the systems. In
Sec.~\ref{sect:indep}, it is proved that these Hamiltonian
systems are actually Liouville integrable for $n=2$. That is,
they still have a full set of involutive and independent conserved
integrals. These conserved integrals are obtained from $\tr
L^k(\lambda)$ and other obvious conserved integrals. When $n>2$,
it is still open whether one can find enough involutive and
independent conserved integrals by adding some obvious ones
to $\tr L^k(\lambda)$'s. Therefore, at least for $n=2$,
the Hamiltonian systems derived from the $U(2)$ principal chiral
field are Liouville integrable, but their conserved integrals for
Liouville integrability can not be fully obtained from $\tr
L^k(\lambda)$.

\section{Hamiltonian systems derived from
\textit{U}(\lowercase{{\large{\textit n}}}) principal chiral
field}\label{sect:hm12}

The equation for the $U(n)$ principal chiral field is
\begin{equation}
   (g_xg^{-1})_t+(g_tg^{-1})_x=0
   \label{eq:pcf_R11org}
\end{equation}
where the field $g(x,t)\in U(n)$. Write
\begin{equation}
   P=g_xg^{-1}\qquad Q=g_tg^{-1}
   \label{eq:PQdef12}
\end{equation}
then $P,Q\in u(n)$ (i.e. $P^*+P=0$, $Q^*+Q=0$) and
(\ref{eq:pcf_R11org}) becomes
\begin{equation}
   P_t+Q_x=0\qquad
   P_t-Q_x+[P,Q]=0.
   \label{eq:pcf_R11}
\end{equation}
Here the second equation is the integrability condition of
(\ref{eq:PQdef12}).

It is known that (\ref{eq:pcf_R11}) has a Lax pair
\begin{equation}
   \Phi_x=\frac 1{1-\lambda}P\Phi\qquad
   \Phi_t=\frac 1{1+\lambda}Q\Phi
   \label{eq:lp_R11}
\end{equation}
where $\lambda$ is a complex spectral parameter.

Now we write down the corresponding finite dimensional
Hamiltonian systems and their Lax operators.

Let $\lambda_1,\cdots,\lambda_N$ be $N$ distinct real constants
with $\lambda_j\ne\pm 1$ $(j=1,\cdots,N)$,
$(\phi_{1\alpha},\cdots,\phi_{n\alpha})^T$ be an arbitrary
solution of the Lax pair (\ref{eq:lp_R11}) with
$\lambda=\lambda_\alpha$,
$\Lambda=\diag(\lambda_1,\cdots,\lambda_N)$,
$\Phi_j=(\phi_{j1},\cdots,\phi_{jN})^T$. Let
\begin{equation}
   L=\sum_{\alpha=1}^N\frac 1{\lambda-\lambda_\alpha}\pmatrix{cccc}
   \bar\phi_{1\alpha}\phi_{1\alpha} &\bar\phi_{2\alpha}\phi_{1\alpha}
    &\cdots &\bar\phi_{n\alpha}\phi_{1\alpha}\\
   \vdots &\vdots &\ddots &\vdots\\
   \bar\phi_{1\alpha}\phi_{n\alpha} &\bar\phi_{2\alpha}\phi_{n\alpha}
    &\cdots &\bar\phi_{n\alpha}\phi_{n\alpha}\endpmatrix.
   \label{eq:L}
\end{equation}
Expand $L$ to power series of $1-\lambda$ and
$1+\lambda$ respectively:
\begin{equation}
   \aligned{l}
   \D L=L^{(1)}=\sum_{k=1}^\infty (1-\lambda)^{k-1}\pmatrix{ccc}
   \langle\Phi_1,(1-\Lambda)^{-k}\Phi_1\rangle
   &\cdots
   &\langle\Phi_n,(1-\Lambda)^{-k}\Phi_1\rangle\\
   \vdots &\ddots &\vdots\\
   \langle\Phi_1,(1-\Lambda)^{-k}\Phi_n\rangle
   &\cdots
   &\langle\Phi_n,(1-\Lambda)^{-k}\Phi_n\rangle
   \endpmatrix\\
   \\
   \D L=L^{(2)}=-\sum_{k=1}^\infty (1+\lambda)^{k-1}\pmatrix{ccc}
   \langle\Phi_1,(1+\Lambda)^{-k}\Phi_1\rangle
   &\cdots
   &\langle\Phi_n,(1+\Lambda)^{-k}\Phi_1\rangle\\
   \vdots &\ddots &\vdots\\
   \langle\Phi_1,(1+\Lambda)^{-k}\Phi_n\rangle
   &\cdots
   &\langle\Phi_n,(1+\Lambda)^{-k}\Phi_n\rangle
   \endpmatrix
   \endaligned
\end{equation}
where the inner product $\langle V_1,V_2\rangle$ of two vectors
is defined as $V_1^*V_2$. The first series converges when
$\D|\lambda-1|<\min_{1\le \alpha\le N}|\lambda_\alpha-1|$ and the
second one converges when $\D|\lambda+1|<\min_{1\le \alpha\le
N}|\lambda_\alpha+1|$.

\begin{lemma}\label{lemma:laxeqR11}
If
\begin{equation}
   P_{jk}=\I\langle\Phi_k,(1-\Lambda)^{-1}\Phi_j\rangle
   \qquad
   Q_{jk}=\I\langle\Phi_k,(1+\Lambda)^{-1}\Phi_j\rangle
   \label{eq:constr_R11}
\end{equation}
then
\begin{equation}
   L_x=\frac 1{1-\lambda}[P,L]\qquad L_t=\frac 1{1+\lambda}[Q,L]
   \label{eq:laxeq1}
\end{equation}
and $(P,Q)$ gives a solution of (\ref{eq:pcf_R11}).
\end{lemma}

\begin{demo}
Let $\phi_\alpha=(\phi_{1\alpha},\cdots,\phi_{n\alpha})^T$, then
\begin{equation}
   L=\sum_{\alpha=1}^N\frac 1{\lambda-\lambda_\alpha}
    \phi_\alpha\phi_\alpha^*.
   \label{eq:LL}
\end{equation}
Since $P^*=-P$ and $\lambda_\alpha$'s are real,
\begin{equation}
   \aligned{rl}
   L_x&\D=\sum_{\alpha=1}^N\frac 1{\lambda-\lambda_\alpha}
    \left(\phi_\alpha\frac 1{1-\lambda_\alpha}\phi_\alpha^*P^*
    +\frac 1{1-\lambda_\alpha}P\phi_\alpha\phi_\alpha^*\right)\\
   &\D=\sum_{\alpha=1}^N\frac 1{\lambda-\lambda_\alpha}
    \frac 1{1-\lambda_\alpha}
    [P,\phi_\alpha\phi_\alpha^*]\\
   &\D=\sum_{\alpha=1}^N\left(\frac 1{\lambda-\lambda_\alpha}
     \frac 1{1-\lambda}
    -\frac 1{1-\lambda_\alpha}\frac 1{1-\lambda}\right)
    [P,\phi_\alpha\phi_\alpha^*]\\
   &\D=\frac 1{1-\lambda}[P,L].
   \endaligned
\end{equation}
The last equality holds due to (\ref{eq:constr_R11}). The equation
for $L_t$ in (\ref{eq:laxeq1}) is derived similarly. Finally, by
computing the integrability condition $L_{xt}=L_{tx}$ from
(\ref{eq:laxeq1}) or substituting (\ref{eq:constr_R11}) into
(\ref{eq:pcf_R11}) directly, we know that $(P,Q)$ satisfies
(\ref{eq:pcf_R11}). The lemma is proved.
\end{demo}

Now we always suppose (\ref{eq:constr_R11}) holds for the $U(n)$
principal chiral field, which gives the nonlinear constraints.
Substituting (\ref{eq:constr_R11}) into (\ref{eq:lp_R11}), we get
a system of partial differential equations
\begin{equation}
   \aligned{l}
   \D\Phi_{j,x}=\I(1-\Lambda)^{-1}\sum_{k=1}^n
    \langle\Phi_k,(1-\Lambda)^{-1}\Phi_j\rangle\Phi_k\\
   \D\Phi_{j,t}=\I(1+\Lambda)^{-1}\sum_{k=1}^n
    \langle\Phi_k,(1+\Lambda)^{-1}\Phi_j\rangle\Phi_k
   \label{eq:ODE_R11}
   \endaligned
\end{equation}
which can studied as two systems of ordinary differential
equations when $t$ and $x$ are considered as constants
respectively.

Now $\phi_{11}, \phi_{12}, \cdots, \phi_{1N}$, $\cdots$,
$\phi_{n1}, \phi_{n2}, \cdots, \phi_{nN}$ and their complex
conjugations form the complex coordinates of $\hr^{2nN}$. In this
$\hr^{2nN}$, let $\omega$ be the standard symplectic form
\begin{equation}
   \aligned{rl}
   \omega&\D=2\sum_{j=1}^n\sum_{\alpha=1}^N
   \d\im(\phi_{j\alpha})\wedge\d\re(\phi_{j\alpha})\\
   &\D=\I\sum_{j=1}^n\sum_{\alpha=1}^N
    \d\bar\phi_{j\alpha}\wedge\d\phi_{j\alpha}
   \endaligned
\end{equation}
then the corresponding Poisson bracket for two functions $f$ and
$g$ is
\begin{equation}
   \{f,g\}=\frac 1\I\sum_{j=1}^n\sum_{\alpha=1}^N\left(
   \frac{\partial f}{\partial\phi_{j\alpha}}
   \frac{\partial g}{\partial\bar\phi_{j\alpha}}
   -\frac{\partial g}{\partial\phi_{j\alpha}}
   \frac{\partial f}{\partial\bar\phi_{j\alpha}}
   \right).
   \label{eq:Poisson}
\end{equation}

From (\ref{eq:laxeq1}), the coefficients of $(1-\lambda)^j$
$(j=0,1,2,\cdots)$ in $\tr(L^{(1)})^k$ $(k=1,2,\cdots)$ and the
coefficients of $(1+\lambda)^j$ $(j=0,1,2,\cdots)$ in
$\tr(L^{(2)})^k$ $(k=1,2,\cdots)$ are all conserved. Suppose
\begin{equation}
   \tr(L^{(1)})^m=\sum_{k=1}^\infty(1-\lambda)^{k-1}\EA1mk\qquad
   \tr(L^{(2)})^m=(-1)^m\sum_{k=1}^\infty(1+\lambda)^{k-1}\EA2mk.
   \label{eq:Lexpand12}
\end{equation}

Since $\tr P=\I\EA111$ and $\tr Q=\I\EA211$, both $\tr P$ and
$\tr Q$ are conserved.
On the other hand, the Hamiltonians for the equations
(\ref{eq:ODE_R11}) are given by $\EA121$ and $\EA221$
according to the following lemma. Moreover, direct computation
shows that they commute with each other under the Poisson
bracket (\ref{eq:Poisson}) (this can also be derived directly
from Lemma~\ref{lemma:invol} in Sec.~\ref{sect:Laxop}).

\begin{lemma}\label{lemma:H12}
The Hamiltonians for the $x$-equation and the $t$-equation of
(\ref{eq:ODE_R11}) are given by
\begin{equation}
   \aligned{l}
   \D H^x=-\frac 12\EA121=-\frac 12\sum_{j,k=1}^n
   \langle\Phi_k,(1-\Lambda)^{-1}\Phi_j\rangle
   \langle\Phi_j,(1-\Lambda)^{-1}\Phi_k\rangle\\
   \D H^t=-\frac 12\EA221=-\frac 12\sum_{j,k=1}^n
   \langle\Phi_k,(1+\Lambda)^{-1}\Phi_j\rangle
   \langle\Phi_j,(1+\Lambda)^{-1}\Phi_k\rangle
   \endaligned
   \label{eq:H12}
\end{equation}
respectively. That is, (\ref{eq:ODE_R11}) is equivalent to the
Hamiltonian equations
\begin{equation}
   \I \phi_{j\alpha,x}=\frac{\partial H^x}{\partial\bar\phi_{j\alpha}}\quad
   -\I\bar\phi_{j\alpha,x}=\frac{\partial H^x}{\partial\phi_{j\alpha}}\quad
   \I \phi_{j\alpha,t}=\frac{\partial H^t}{\partial\bar\phi_{j\alpha}}\quad
   -\I\bar\phi_{j\alpha,t}=\frac{\partial H^t}{\partial\phi_{j\alpha}}.
\end{equation}
Moreover, $\{H^x,H^t\}=0$.
\end{lemma}

\begin{remark}
The above procedure can also be used for the harmonic map from
$\hr^2$ to $U(n)$. In this case, the equation is
\begin{equation}
   (g_zg^{-1})_{\bar z}+(g_{\bar z}g^{-1})_z=0
\end{equation}
where $z$ is the complex coordinate of $\hr^2$, $g(z,\bar z)\in U(n)$.
The Lax pair is
\begin{equation}
   \Phi_z=\frac 1{1-\I\lambda}g_zg^{-1}\Phi\qquad
   \Phi_{\bar z}=\frac 1{1+\I\lambda}g_{\bar z}g^{-1}\Phi.
\end{equation}
where $\lambda$ is a complex spectral parameter. Using the same
method in the last section, we can also get finite dimensional
Hamiltonian systems whose Lax operator is completely the same as
(\ref{eq:L}).
\end{remark}

\section{Conserved integrals}\label{sect:Laxop}

\begin{lemma}\label{lemma:invol}
With the Poisson bracket (\ref{eq:Poisson}), the following two
conclusions hold.

(1) For any two complex numbers $\lambda$, $\mu$ and two positive
integers $k$, $l$,
\begin{equation}
   \{\tr L^k(\lambda),\tr L^l(\mu)\}=0.
\end{equation}

(2) For any complex number $\lambda$ and integers $j$, $k$, $l$ with
$1\le j,k\le n$,
\begin{equation}
   \{\langle\Phi_j,\Phi_k\rangle,\tr L^l(\lambda)\}=0.
\end{equation}
\end{lemma}

This can be verified by direct computation of the Poisson brackets
and was given in Ref.~\onlinecite{bib:ZMZ}.

Suppose the eigenvalues of $L(\lambda)$ are $\nu_1(\lambda)$,
$\nu_2(\lambda)$, $\cdots$, $\nu_n(\lambda)$, then
\begin{equation}
   \aligned{l}
   \tr L^k(\lambda)=\nu_1^k(\lambda)+\cdots+\nu_n^k(\lambda)\qquad
   (k=1,2,\cdots)\\
   \det (\mu-L(\lambda))=\mu^n-p_1(\lambda)\mu^{n-1}+\cdots+(-1)^np_n(\lambda)
   \endaligned
\end{equation}
for any complex number $\mu$ where
\begin{equation}
   p_k(\lambda)=\sum_{1\le j_1<\cdots<j_k\le n}
   \nu_{j_1}(\lambda)\cdots\nu_{j_k}(\lambda)
\end{equation}
is the sum of all the determinants of the principal submatrices
of $L(\lambda)$ of order $k$. Hence $\tr L^k(\lambda)$
$(k=1,2,\cdots)$ are uniquely determined by $p_k(\lambda)$
$(k=1,2,\cdots,n)$ and vise versa. Moreover, $\tr(L^{(1)})^k$ and
$\tr(L^{(2)})^k$ in (\ref{eq:Lexpand12}) can all
be uniquely determined by $p_k(\lambda)$ $(k=1,2,\cdots,n)$.

Each $p_k(\lambda)$ is a holomorphic function of $\lambda$ near
$\lambda=\infty$. Let
\begin{equation}
   p_m(\lambda)=\sum_{k=0}^\infty\EE mk\lambda^{-k-m}
\end{equation}
then
\begin{equation}
   \aligned{l}
   \EE mk\\
   \D=\sum_{\scriptstyle 1\le i_1<\cdots<i_m\le n}
   \sum_{\scriptstyle r_1+\cdots+r_m=k\atop
    \scriptstyle r_1,\cdots,r_m\ge 0}
   \left|\matrix{cccc}
   \langle\Phi_{i_1},\Lambda^{r_1}\Phi_{i_1}\rangle
    &\langle\Phi_{i_2},\Lambda^{r_2}\Phi_{i_1}\rangle
    &\cdots
    &\langle\Phi_{i_m},\Lambda^{r_m}\Phi_{i_1}\rangle\\
   \langle\Phi_{i_1},\Lambda^{r_1}\Phi_{i_2}\rangle
    &\langle\Phi_{i_2},\Lambda^{r_2}\Phi_{i_2}\rangle
    &\cdots
    &\langle\Phi_{i_m},\Lambda^{r_m}\Phi_{i_2}\rangle\\
   \vdots &\vdots &\ddots &\vdots\\
   \langle\Phi_{i_1},\Lambda^{r_1}\Phi_{i_m}\rangle
    &\langle\Phi_{i_2},\Lambda^{r_2}\Phi_{i_m}\rangle
    &\cdots
    &\langle\Phi_{i_m},\Lambda^{r_m}\Phi_{i_m}\rangle
   \endmatrix\right|\HHH{36}\\
   \D=\sum_{1\le i_1<\cdots<i_m\le n}
   \sum_{\scriptstyle r_1+\cdots+r_m=k\atop
    \scriptstyle r_1,\cdots,r_m\ge 0}
   \sum_{\scriptstyle \alpha_1,\cdots,\alpha_m=1\atop
    \scriptstyle \alpha_a\ne\alpha_b \hbox{\scriptrm\ for }a\ne b}^N
   \lambda_{\alpha_1}^{r_1}\cdots\lambda_{\alpha_m}^{r_m}\\
   \qquad\times\left|\matrix{cccc}
   \bar\phi_{i_1\alpha_1}\phi_{i_1\alpha_1}
    &\bar\phi_{i_2\alpha_2}\phi_{i_1\alpha_2}
    &\cdots
    &\bar\phi_{i_m\alpha_m}\phi_{i_1\alpha_m}\\
   \bar\phi_{i_1\alpha_1}\phi_{i_2\alpha_1}
    &\bar\phi_{i_2\alpha_2}\phi_{i_2\alpha_2}
    &\cdots
    &\bar\phi_{i_m\alpha_m}\phi_{i_2\alpha_m}\\
   \vdots &\vdots &\ddots &\vdots\\
   \bar\phi_{i_1\alpha_1}\phi_{i_m\alpha_1}
    &\bar\phi_{i_2\alpha_2}\phi_{i_m\alpha_2}
    &\cdots
    &\bar\phi_{i_m\alpha_m}\phi_{i_m\alpha_m}
   \endmatrix\right|.\HHH{36}
   \endaligned
   \label{eq:energy}
\end{equation}
In the last summation, the condition ``$\alpha_a\ne\alpha_b$ for
$a\ne b$'' is added since the determinants with
$\alpha_a=\alpha_b$ $(a\ne b)$ are all zero.

\begin{remark}\label{remark:empty}
When $m\ge N+1$, the last summation in (\ref{eq:energy}) for
``$1\le\alpha_1,\cdots,\alpha_m\le N$ with $\alpha_a\ne \alpha_b$
for $a\ne b$'' is empty. This means that $\EE mk\equiv 0$ for
$m\ge N+1$.
\end{remark}

According to (\ref{eq:laxeq1}), all $\EE
mk$'s are conserved.

From the first part of Lemma~\ref{lemma:invol}, all $\EE mk$'s
are in involution. The second part of Lemma~\ref{lemma:invol}
implies that all $\langle\Phi_j,\Phi_k\rangle$'s commute with
$\EE mk$'s. However, these $\langle\Phi_j,\Phi_k\rangle$'s may
not commute with each other.

\begin{remark}
For the Heisenberg ferromagnetic equation, the $x$-equation of its
Lax pair is similar to that of the $U(2)$ principal chiral field.
The nonlinearization for this equation was dealt with
in Ref.~\onlinecite{bib:Caomag} and a set of involutive conserved integrals
were obtained there.
\end{remark}

\begin{remark}
Since $\tr P$, $\tr Q$ are conserved, if $(P,Q)$ is a solution of
(\ref{eq:pcf_R11}) in $u(n)$, then
\begin{equation}
   P'=P-\frac 1n\tr P\qquad Q'=Q-\frac 1n\tr Q
   \label{eq:su}
\end{equation}
gives a solution of the same equation (\ref{eq:pcf_R11}) in $su(n)$.
\end{remark}

\section{Dependence of conserved integrals}\label{sect:depend}

In order to consider the integrability of the Hamiltonian
systems, we should find a full set of involutive and independent
conserved integrals. Unlike many other cases, here we cannot get
a full set of independent conserved integrals simply from $\tr
L^k(\lambda)$'s.

For further discussion, we need the following lemma.

\begin{lemma}\label{lemma:alg}
Suppose $k$, $m$ are two integers with $k\ge 0$ and $m\ge 2$,
$\mu_1$, $\cdots$, $\mu_m$ are distinct complex numbers,
then
\begin{equation}
   \sum_{j=1}^m\mu_j^k
   \prod_{\scriptstyle r=1\atop\scriptstyle r\ne j}^m
   (\mu_j-\mu_r)^{-1}
   =\left\{\begin{array}{ll} 0 &\hbox{\rm if }k<m-1\\
    \D\sum_{\scriptstyle p_1+\cdots+p_m=k-m+1\atop\scriptstyle
     p_1,\cdots,p_m\ge 0}
   \mu_1^{p_1}\cdots\mu_m^{p_m}\quad &\hbox{\rm if }k\ge m-1.
   \end{array}\right. 
\end{equation}
\end{lemma}

\begin{demo}
Let
\begin{equation}
   f(\zeta)=\zeta^k\prod_{r=1}^m(\zeta-\mu_r)^{-1}.
\end{equation}
$f(\zeta)$ is a meromorphic function of $\zeta$ with poles $\mu_1$,
$\cdots$, $\mu_m$. Let $C_R$ be a circle with radius $R$, center
$0$ and positive orientation, then, when $\D R>\max_{1\le j\le
m}|\mu_j|$,
\begin{equation}
   \frac 1{2\pi\I}\int_{C_R}f(\zeta)\,\d\zeta
   =\sum_{j=1}^m\res_{\zeta=\mu_j}f(\zeta)
   =\sum_{j=1}^m\mu_j^k
   \prod_{\scriptstyle r=1\atop\scriptstyle r\ne j}^m(\mu_j-\mu_r)^{-1}.
\end{equation}
On the other hand, let $\xi=\zeta^{-1}$, then
\begin{equation}
   \aligned{l}
   \D\frac 1{2\pi\I}\int_{C_R}f(\zeta)\,\d\zeta
   =\lim_{R\to\infty}\frac 1{2\pi\I}\int_{C_{1/R}}
   \xi^{m-k-2}\prod_{r=1}^m(1-\mu_r\xi)^{-1}\,\d\xi\\ \\
   =\left\{\begin{array}{ll} 0 &\hbox{\rm if }k-m+1<0\\
   \D\sum_{\scriptstyle p_1+\cdots+p_m=k-m+1\atop\scriptstyle
     p_1,\cdots,p_m\ge 0}
   \mu_1^{p_1}\cdots\mu_m^{p_m}\quad &\hbox{\rm if }k-m+1\ge 0
   \end{array}\right. 
   \endaligned
\end{equation}
by expanding all the terms $(1-\mu_j\xi)^{-1}$ at $\xi=0$. The
lemma is proved.
\end{demo}

\begin{theorem}\label{thm:depend}
For $1\le m\le n$, there are at most $\max(0,N-m+1)$ linearly
independent functions in $\EE mk$ $(k=0,1,2,\cdots)$. Therefore,
the number of linearly independent functions in $\EE mk$
$(m=1,2,\cdots,n;\,k=0,1,2,\cdots)$ cannot exceed $\D nN-\frac
12n(n-1)$ if $N\ge n$ or $\D\frac 12N(N+1)$ if $N<n$.
\end{theorem}

\begin{demo}
According to Remark~\ref{remark:empty}, $\EE mk\equiv 0$ for
$m\ge N+1$. Hence we always suppose $m\le N$.

By definition, $\Lambda=\diag(\lambda_1,\cdots,\lambda_N)$,
$\lambda_j\ne \lambda_k$ $(j\ne k)$. Clearly, for any
non-negative integers $(k_1,\cdots,k_l)$ with $k_i\ne k_j$
$(i\ne j)$ and $l\ge N+1$, $\EE1{k_1},\cdots,\EE1{k_l}$ are
linearly dependent. On the other hand, since the Van de Monde
determinant of $\lambda_1,\cdots,\lambda_N$ is not zero, there
are exactly $N$ independent functions in $\EE 1k$
$(k=0,1,2,\cdots)$.

For $m\ge 2$, we show that there are at most $N-m+1$ independent
functions in $\EE mk$ $(k=0,1,2,\cdots)$ for fixed $m$.

Let $k_1,\cdots,k_N$ be $N$ arbitrary distinct non-negative
integers. For fixed $s$ with $0\le s\le m-2$, let
$(\gamma_1^{(s)},\cdots,\gamma_N^{(s)})$ be
a solution of the linear algebraic system
\begin{equation}
   \sum_{j=1}^N\lambda_\alpha^{k_j+m-1}\gamma_j^{(s)}=\lambda_\alpha^s
   \qquad
   (\alpha=1,2,\cdots,N).
   \label{eq:gamma}
\end{equation}
Since the coefficient matrix in (\ref{eq:gamma}) is
$(\lambda_{\alpha}^{k_j+m+1})_{\alpha,j=1,\cdots,N}$ which is
invertible, $(\gamma_1^{(s)},\cdots,\gamma_N^{(s)})$
exists uniquely.

Let
\begin{equation}
   \EF ms=\sum_{j=1}^N\gamma_j^{(s)}\EE m{k_j}\qquad
   (s=0,1,\cdots,m-2).
\end{equation}
Then
\begin{equation}
   \aligned{l}
   \D \EF ms=\sum_{j=1}^N\sum_{1\le i_1<\cdots<i_m\le n}
   \sum_{\scriptstyle r_1+\cdots+r_m=k_j\atop
    \scriptstyle r_1,\cdots,r_m\ge 0}
   \sum_{\scriptstyle \alpha_1,\cdots,\alpha_m=1\atop
    \scriptstyle \alpha_a\ne\alpha_b \hbox{\scriptrm\ for }a\ne b}^N
   \gamma_j^{(s)}\lambda_{\alpha_1}^{r_1}\cdots\lambda_{\alpha_m}^{r_m}\\
   \qquad\times\left|\matrix{cccc}
   \bar\phi_{i_1\alpha_1}\phi_{i_1\alpha_1}
    &\bar\phi_{i_2\alpha_2}\phi_{i_1\alpha_2}
    &\cdots
    &\bar\phi_{i_m\alpha_m}\phi_{i_1\alpha_m}\\
   \bar\phi_{i_1\alpha_1}\phi_{i_2\alpha_1}
    &\bar\phi_{i_2\alpha_2}\phi_{i_2\alpha_2}
    &\cdots
    &\bar\phi_{i_m\alpha_m}\phi_{i_2\alpha_m}\\
   \vdots &\vdots &\ddots &\vdots\\
   \bar\phi_{i_1\alpha_1}\phi_{i_m\alpha_1}
    &\bar\phi_{i_2\alpha_2}\phi_{i_m\alpha_2}
    &\cdots
    &\bar\phi_{i_m\alpha_m}\phi_{i_m\alpha_m}
   \endmatrix\right|.\HHH{36}
   \endaligned
\end{equation}
For fixed $i_1,\cdots,i_m$, $\alpha_1,\cdots,\alpha_m$ and $s$
with $0\le s\le m-2$, let
\begin{equation}
   \Delta=\sum_{j=1}^N
   \sum_{\scriptstyle r_1+\cdots+r_m=k_j\atop
    \scriptstyle r_1,\cdots,r_m\ge 0}
   \gamma_j^{(s)}\lambda_{\alpha_1}^{r_1}\cdots\lambda_{\alpha_m}^{r_m}
\end{equation}
then
\begin{equation}
   \Delta=\sum_{j=1}^N\gamma_j^{(s)}\sum_{a=1}^m\lambda_{\alpha_a}^{k_j+m-1}
   \prod_{\scriptstyle r=1\atop\scriptstyle r\ne a}^m
    (\lambda_{\alpha_a}-\lambda_{\alpha_r})^{-1}
\end{equation}
by Lemma~\ref{lemma:alg}. The relations (\ref{eq:gamma}) imply
\begin{equation}
   \Delta=\sum_{a=1}^m\lambda_{\alpha_a}^s
   \prod_{\scriptstyle r=1\atop\scriptstyle r\ne a}^{m}
    (\lambda_{\alpha_a}-\lambda_{\alpha_r})^{-1}.
\end{equation}
Using Lemma~\ref{lemma:alg} again, we get $\Delta=0$ for
$s=0,1,2,\cdots,m-2$. Hence
\begin{equation}
   \EF m0=\EF m1=\cdots=\EF m{m-2}=0.
\end{equation}
By (\ref{eq:gamma}), the matrix $(\gamma_j^{(s)})_{1\le j\le N;\,0\le
s\le m-2}$ has rank $m-1$. Hence $\EE m{k_j}$ $(j=1,2,\cdots,N;$
$m=0,1,\cdots,m-2)$ satisfy $m-1$ independent linear relations for
fixed $m$. This means that there are at most $N-m+1$ independent
functions in $N$ functions $\EE m{k_j}$ for fixed $m$.

Since $k_1,\cdots,k_N$ are arbitrary, there are at most $N-m+1$
independent functions in $\EE mk$ $(k=0,1,2,\cdots)$.

The total number of possible independent functions in $\EE mk$
$(m=1,2,\cdots,n;$ $k=0,1,2,\cdots)$ is
\begin{equation}
   \sum_{m=1}^n(N-m+1)=nN-\frac 12n(n-1)
\end{equation}
for $N\ge n$ and
\begin{equation}
   \sum_{m=1}^N(N-m+1)=\frac 12N(N+1)
\end{equation}
for $N<n$. The theorem is proved.
\end{demo}

A completely integrable Hamiltonian system in $\hr^{2nN}$ needs
$nN$ independent involutive conserved integrals. Hence the above
theorem shows that it is not possible to find enough conserved
integrals only from $\EE mk$'s for Liouville integrability.

\section{Liouville integrability of the Hamiltonian systems}
\label{sect:indep}

In general, we have not been able to determine whether the
Hamiltonian systems for the $U(n)$ principal chiral field are
Liouville integrable or not. However, when $n=2$, the answer is
positive.

Hereafter, we suppose $n=2$. Therefore, we want to find $2N$
independent conserved integrals for the Hamiltonian systems in
$\hr^{4N}$.

If $N=1$, let
\begin{equation}
   \aligned{l}
   \EEE 10=\EE 10=\langle\Phi_1,\Phi_1\rangle+\langle\Phi_2,\Phi_2\rangle\\
   \EEE 20=\langle\Phi_1,\Phi_2\rangle+\langle\Phi_2,\Phi_1\rangle.
   \endaligned
\end{equation}
If $N\ge 2$, let
\begin{equation}
   \aligned{l}
   \EEE 1k=\EE 1k=\langle\Phi_1,\Lambda^k\Phi_1\rangle
    +\langle\Phi_2,\Lambda^k\Phi_2\rangle
    \qquad (k=0,1,\cdots,N-1)\\
   \D\EEE 2k=\EE 2k=\sum_{j=0}^k\left|\matrix{cc}
    \langle\Phi_1,\Lambda^j\Phi_1\rangle
    &\langle\Phi_2,\Lambda^{k-j}\Phi_1\rangle\\
    \langle\Phi_1,\Lambda^j\Phi_2\rangle
    &\langle\Phi_2,\Lambda^{k-j}\Phi_2\rangle\endmatrix\right|
    \qquad (k=0,1,\cdots,N-2)\\
   \EEE 2{N-1}=\langle\Phi_1,\Phi_2\rangle+\langle\Phi_2,\Phi_1\rangle.
   \label{eq:intn=2}
   \endaligned
\end{equation}
Here the last one is chosen to be
$\langle\Phi_1,\Phi_2\rangle+\langle\Phi_2,\Phi_1\rangle$
because all the conserved integrals should take real value.

\begin{theorem}\label{thm:indep}
When $n=2$, $\EEE mk$ $(m=1,2;\,k=0,1,\cdots,N-1)$ are in
involution and are functionally independent in a dense open
subset of $\hr^{4N}$.
\end{theorem}

\begin{demo}
By Lemma~\ref{lemma:invol}, $\EEE mk$ $(m=1,2;\,k=0,1,\cdots,N-1)$
are in involution.

It is obvious that they are independent for $N=1$. Hence we
suppose $N\ge 2$. Let $a_{1\alpha}$ $(\alpha=1,2,\cdots,N)$ be
$N$ non-zero real numbers,
\begin{equation}
   a_{2\alpha}=a_{1\alpha}^{-1}
   \prod_{\scriptstyle \beta=1\atop\scriptstyle \beta\ne \alpha}^N
   (\lambda_\alpha-\lambda_\beta)^{-1}
   \qquad (\alpha=1,2,\cdots,N).
   \label{eq:a12}
\end{equation}
Then Lemma~\ref{lemma:alg} implies
\begin{equation}
   \aligned{l}
   \D\sum_{\beta=1}^N\lambda_\beta^k\bar a_{2\beta}a_{1\beta}=0
   \qquad (k=0,1,\cdots,N-2)\\
   \D\sum_{\beta=1}^N\lambda_\beta^{N-1}\bar a_{2\beta}a_{1\beta}=1.
   \endaligned \label{eq:orthpt}
\end{equation}

Let $P_0\in\hr^{4N}$ be given by $\phi_{1\beta}=a_{1\beta}$,
$\phi_{2\beta}=\epsilon a_{2\beta}$ $(\beta=1,2,\cdots,N)$. Here
$\epsilon$ is a non-zero small constant to be determined. Then,
at $P_0$,
\begin{equation}
   \aligned{l}
   \D\frac{\partial\EEE 1k}{\partial\bar\phi_{1\alpha}}
    =\lambda_\alpha^k\phi_{1\alpha}\qquad
   \frac{\partial\EEE 1k}{\partial\bar\phi_{2\alpha}}
    =\lambda_\alpha^k\phi_{2\alpha}\HHH{20}\\
   \qquad\qquad (k=0,1,\cdots,N-1)\\
   \D\frac{\partial\EEE 2k}{\partial\bar\phi_{1\alpha}}
    =\sum_{j=0}^k\left|\matrix{cc}
    \lambda_\alpha^j\phi_{1\alpha}\quad
    &\D\sum_{\beta=1}^N\lambda_{\beta}^{k-j}\bar\phi_{2\beta}
    \phi_{1\beta}\\
    \lambda_\alpha^j\phi_{2\alpha}\quad
    &\D\sum_{\beta=1}^N\lambda_{\beta}^{k-j}|\phi_{2\beta}|^2
   \endmatrix\right|
   =\sum_{j=0}^k r_{2,k-j}
    \lambda_\alpha^j\phi_{1\alpha}\HHH{40}\\
   \D\frac{\partial\EEE 2k}{\partial\bar\phi_{2\alpha}}
    =\sum_{j=0}^k\left|\matrix{cc}
    \D\sum_{\beta=1}^N\lambda_{\beta}^{k-j}|\phi_{1\beta}|^2
    &\quad\lambda_\alpha^j\phi_{1\alpha}\\
    \D\sum_{\beta=1}^N\lambda_{\beta}^{k-j}\bar\phi_{1\beta}\phi_{2\beta}
    &\quad\lambda_\alpha^j\phi_{2\alpha}
   \endmatrix\right|
   =\sum_{j=0}^k r_{1,k-j}
    \lambda_\alpha^j\phi_{2\alpha}\HHH{42}\\
   \qquad\qquad(k=0,1,\cdots,N-2)\HHH{16}\\
   \D\frac{\partial\EEE 2{N-1}}{\partial\bar\phi_{1\alpha}}
   =\phi_{2\alpha}\qquad
   \frac{\partial\EEE 2{N-1}}{\partial\bar\phi_{2\alpha}}
   =\phi_{1\alpha}\HHH{20}
  \endaligned
\end{equation}
by using (\ref{eq:orthpt}) where
\begin{equation}
   r_{jk}=\sum_{\beta=1}^N\lambda_\beta^k|\phi_{j\beta}|^2.
\end{equation}

Let $J$ be the Jacobian matrix
\begin{equation}
   J=\left.\frac{\partial(\EEE 10,\cdots,\EEE 1{N-1},\EEE 20,\cdots,
   \EEE 2{N-1})}
   {\partial(\bar \phi_{11},\cdots,
    \bar \phi_{1N},\bar \phi_{21},\cdots,\bar\phi_{2N})}\right|_{P_0}.
\end{equation}

\def\ROW#1{\hbox{\ss ROW}{}_{#1}}
Denote $\ROW{j}$ to be the $j$-th row of $J$. Take the elementary
transformations for the rows of $J$ as follows:
\begin{equation}
   \aligned{l}
   \hbox{(1) } k\hbox{ from }1\hbox{ to }N-1:\\
   \qquad\D \ROW{N+k}-\sum_{j=0}^{k-1}r_{2,k-1-j}\ROW{j+1}\to\ROW{N+k}\\
   \hbox{(2) } k\hbox{ from }2\hbox{ to }N-1:\\
   \qquad\D\ROW{N+k}-\sum_{j=1}^{k-1}\frac{r_{1j}-r_{2j}}{r_{10}-r_{20}}
    \ROW{N+k-j}\to\ROW{N+k}\\
   \hbox{(3) } k\hbox{ from }1\hbox{ to }N-1:\\
   \qquad (r_{10}-r_{20})^{-1}\ROW{N+k}\to \ROW{N+k}\\
   \hbox{(4) } k\hbox{ from }1\hbox{ to }N-1:\\
   \qquad \ROW{k}-\ROW{N+k}\to \ROW{k}
   \endaligned\nonumber
\end{equation}
then $J$ is transformed to
\begin{equation}
   \widetilde J=\pmatrix{cccccccc}
    \phi_{11} &\phi_{12} &\cdots &\phi_{1N} &0 &0 &\cdots &0\\
    \lambda_1\phi_{11} &\lambda_2\phi_{12} &\cdots &\lambda_N\phi_{1N}
     &0 &0 &\cdots &0\\
    \vdots &\vdots &\ddots &\vdots &\vdots &\vdots &\ddots &\vdots\\
    \lambda_1^{N-2}\phi_{11} &\lambda_2^{N-2}\phi_{12} &\cdots
     &\lambda_N^{N-2}\phi_{1,N-1}
     &0 &0 &\cdots &0\\
    \lambda_1^{N-1}\phi_{11} &\lambda_2^{N-1}\phi_{12} &\cdots
     &\lambda_N^{N-1}\phi_{1N}
     &\lambda_1^{N-1}\phi_{21} &\lambda_2^{N-1}\phi_{22} &\cdots
     &\lambda_N^{N-1}\phi_{2N}\\
    0 &0 &\cdots &0 &\phi_{21} &\phi_{22} &\cdots &\phi_{2N}\\
    0 &0 &\cdots &0
     &\lambda_1\phi_{21} &\lambda_2\phi_{22} &\cdots &\lambda_N\phi_{2N}\\
    \vdots &\vdots &\ddots &\vdots &\vdots &\vdots &\ddots &\vdots\\
    0 &0 &\cdots &0
     &\lambda_1^{N-2}\phi_{21} &\lambda_2^{N-2}\phi_{22} &\cdots
     &\lambda_N^{N-2}\phi_{2N}\\
    \phi_{21} &\phi_{22} &\cdots &\phi_{2N}
    &\phi_{11} &\phi_{12} &\cdots &\phi_{1N}.
   \endpmatrix.
\end{equation}
Let
\begin{equation}
   T=\pmatrix{cc} T_2 &\\&T_1\endpmatrix
\end{equation}
where
\begin{equation}
   T_j=\left.\pmatrix{cccc}
    \lambda_1^{N-1}\bar \phi_{j1} &\lambda_1^{N-2}\bar\phi_{j1} &\cdots
    &\bar \phi_{j1}\\
    \lambda_2^{N-1}\bar \phi_{j2} &\lambda_2^{N-2}\bar\phi_{j2} &\cdots
    &\bar \phi_{j2}\\
    \vdots &\vdots &\ddots &\vdots\\
    \lambda_N^{N-1}\bar \phi_{jN} &\lambda_N^{N-2}\bar\phi_{jN} &\cdots
    &\bar \phi_{jN}
   \endpmatrix\right|_{P_0}\qquad (j=1,2)
\end{equation}
then
\begin{equation}
   \det T=\prod_{1\le\alpha<\beta\le N}
   (\lambda_\alpha-\lambda_\beta)^2\prod_{\gamma=1}^N
   \bar\phi_{1\gamma}\bar\phi_{2\gamma}|_{P_0}\ne 0.
\end{equation}
Using the relations
(\ref{eq:orthpt}), we have, at $P_0$,
\begin{equation}
   \widetilde JT=\pmatrix{ccccccccccc}
   &\rho &0 &\cdots &0 &0 &0 &0 &\cdots &0 &0\\
   &* &\rho &\cdots &0 &0 &0 &0 &\cdots &0 &0\\
   &\vdots &\vdots & &\vdots &\vdots &\vdots &\vdots & &\vdots &\vdots\\
   &* &* &\cdots &\rho &0 &0 &0 &\cdots &0 &0\\
   &* &* &\cdots &* &\rho &* &* &\cdots &* &\bar\rho\\
   &0 &0 &\cdots &0 &0 &\rho &0 &\cdots &0 &0\\
   &0 &0 &\cdots &0 &0 &* &\rho &\cdots &0 &0\\
   &\vdots &\vdots & &\vdots &\vdots &\vdots &\vdots & &\vdots &\vdots\\
   &0 &0 &\cdots &0 &0 &* &* &\cdots &\rho &0\\
   &* &* &\cdots &* &\D\sum_{j=1}^N|\phi_{2j}|^2
   &* &* &\cdots &* &\D\sum_{j=1}^N|\phi_{1j}|^2
   \endpmatrix
\end{equation}
where
\begin{equation}
   \rho=\left.\sum_{\beta=1}^N\lambda_\beta^{N-1}
    \bar\phi_{2\beta}\phi_{1\beta}\right|_{P_0}=\epsilon\ne 0
\end{equation}
and $*$ represents the entries which may not be zero.

Hence, at $P_0$,
\begin{equation}
   \aligned{l}
   \D(r_{10}-r_{20})^{-N+1}\det(JT)=\det(\widetilde JT)=\rho^{2N-2}\left(
    \rho\sum_{j=1}^N|\phi_{1j}|^2
    -\bar\rho\sum_{j=1}^N|\phi_{2j}|^2\right)\\
   \D=\epsilon^{2N-1}
   \left(
    \sum_{j=1}^N a_{1j}^2
    -\epsilon^2\sum_{j=1}^N a_{2j}^2\right).
   \endaligned
\end{equation}
It is not zero when $\epsilon$ is small enough. Since $\det J$ is a
real analytical function on $\hr^{4N}$, $\det J$ is not zero in a
dense open subset of $\hr^{4N}$. The theorem is proved.
\end{demo}

\begin{remark}
Although the constraint here is of Bargmann type, the proof of the
in\-de\-pen\-dence of the conserved integrals is not so simple as in the
AKNS system. In that case, $P_0$ is simply chosen as a point near
$0$. However, here $L(\lambda)$ is homogeneous to all $\Phi_j$'s
so the choice of $P_0$ near $0$ has not any effect in
simplification of the computation on $J$.
\end{remark}

The Liouville integrability of the $U(2)$ principal chiral field
follows from Lemma~\ref{lemma:H12} and Theorem~\ref{thm:indep}.
It is given by the following theorem.

\begin{theorem}
When $n=2$, the Hamiltonian systems given by (\ref{eq:H12}) are
completely integrable in Liouville sense. Each
solution of the Hamiltonian systems (\ref{eq:H12}) gives a
solution $(P,Q)$ of (\ref{eq:pcf_R11}), the equation of $U(2)$
principal chiral field, and $(P-\frac 12\tr P,Q-\frac 12\tr Q)$ is a
solution of $SU(2)$ principal chiral field.
\end{theorem}

\begin{remark}
Theorem~\ref{thm:depend} implies that one needs at least
$n(n+1)/2$ extra conserved integrals together with $\EE mk$'s to
form a full set of conserved integrals for the complete
integrability of the Hamiltonian systems. According to
Lemma~\ref{lemma:invol}, all $\langle\Phi_k,\Phi_j\rangle$'s
commute with $\EE mk$.  However, two elements in
$\{\langle\Phi_k,\Phi_j\rangle\}$ may not commute with each
other. Therefore, it is not obvious how to add at least
$n(n+1)/2$ extra conserved integrals to $\EE mk$ in general.
\end{remark}

\section*{Acknowledgements}
This work was supported by the Special Funds for Chinese Major
State Basic Research Projects, the Doctoral Program Foundation,
the Trans-century Training Program Foundation for the Talents and
the Foundation for University Key Teacher by the Ministry of
Education of China. The author is  grateful to Prof.~C.~H.~Gu and
Prof.~H.~S.~Hu for their suggestions and Dr.~R.~G.~Zhou and
Dr.~W.~X.~Ma for helpful discussions.

\end{document}